\documentclass[useAMS,usenatbib,usegraphicx]{mn2e}

\title[First e-VLBI observations of Cygnus X-3]{First e-VLBI observations of Cygnus X-3}
\author[V. Tudose et al.]{V. Tudose,$^{1,2}$\thanks{E-mail: vtudose@science.uva.nl (VT)} R.P. Fender,$^{3,1}$ M.A. Garrett,$^{4}$ 
J.C.A. Miller-Jones,$^{1}$ Z. Paragi,$^{4}$  
\newauthor  R.E. Spencer,$^{5}$ G.G. Pooley,$^{6}$ M. van der Klis$^{1}$ and A. Szomoru$^4$\\
$^{1}$Astronomical Institute `Anton Pannekoek', University of Amsterdam, Kruislaan 403, 1098 SJ Amsterdam, The Netherlands\\
$^{2}$Astronomical Institute of the Romanian Academy, Cutitul de Argint 5, RO-040557 Bucharest, Romania\\
$^{3}$School of Physics and Astronomy, University of Southampton, Highfield, Southampton SO17 1BJ\\
$^{4}$Joint Institute for VLBI in Europe, Postbus 2, 7990 AA Dwingeloo, The Netherlands\\
$^{5}$University of Manchester, Jodrell Bank Observatory, Macclesfield, Cheshire SK11 9DL\\
$^{6}$University of Cambridge, Mullard Radio Astronomy Observatory, J.J. Thomson Avenue, Cambridge CB3 0HE}

\begin{document}

\date{Accepted XXXXXX. Received XXXXXX; in original form XXXXXX}

\pagerange{\pageref{firstpage}--\pageref{lastpage}} \pubyear{2006}

\maketitle

\label{firstpage}

\begin{abstract}

We report the results of the first two 5~GHz e-VLBI observations of
the X-ray binary Cygnus X-3 using the European VLBI Network.  Two
successful observing sessions were held, on 2006 April 20, when the
system was in a quasi-quiescent state several weeks after a major
flare, and on 2006 May 18, a few days after another flare. At the
first epoch we detected faint emission probably associated with a
fading jet, spatially separated from the X-ray binary. The second
epoch in contrast reveals a bright, curved, relativistic jet more than
40 milliarcseconds in extent. In the first, and probably also second
epochs, the X-ray binary core is not detected, which may indicate a
temporary suppression of jet production as seen in some black hole
X-ray binaries in certain X-ray states. Spatially resolved
polarisation maps at the second epoch provide evidence of interaction
between the ejecta and the surrounding medium. These results
clearly demonstrate the importance of rapid analysis of long-baseline
observations of transients, such as facilitated by e-VLBI.

\end{abstract}


\section{Introduction}

The X-ray binary Cygnus X-3 was first detected in X-rays by
\cite{Gia67}. The infrared (e.g.\ \citealt{Bec73}) and X-ray fluxes
(e.g.\ \citealt{Par72}) show a periodicity of 4.8 hours which is
interpreted as the orbital period of the system. The nature of the
compact object is not known \citep*{Sch96,Mit98}.  As for the
companion star, there is compelling evidence pointing toward a WN 
Wolf-Rayet star \citep*{Ker96,Fen99,Koc02}.

Giant outbursts and large flares have been observed at radio
wavelengths in Cygnus X-3 since 1972 \citep{Gre72}. In quiescence the
soft X-ray emission is correlated with the radio emission, while the hard X-ray
is anti-correlated with the radio; in a flare state, the 
situation is reversed: the hard X-ray correlates with the radio and 
the soft X-ray emission is anti-correlated \citep{Wat94,Col99,Cho02}.

Radio observations made during such large flares at different
resolutions with the Very Large Array (VLA), Multi-Element Radio Linked 
Interferometer Network (MERLIN), Very Long Baseline
Array (VLBA), and European VLBI Network (EVN)
\citep*{Gel83,Spe86,Mol88,Sch95,Sch98,Mio01,Mar01,Mil04} directly show
or are consistent with two-sided relativistic jets (with the notable
exception of the VLBA observations of a flare in February 1997, when
the jet was apparently one-sided; \citet{Mio01}).

\section{Observations}

One of the aims of e-VLBI is to enable mapping with long-baseline 
networks of radio telescopes in a manner which makes it possible 
to map transient phenomena, such as microquasars, in near real-time.
This will provide the ability to make informed decisions about the
optimum observing strategy to employ (frequency of observations, 
array composition, calibrator selection, etc.) and the need for repeated 
mapping observations, as well as greatly simplifying 
the observing procedure and improving its reliability. In the case of the EVN, 
the data are transported 
to the correlator at JIVE (Joint Institute for VLBI in Europe) in real-time using 
IP routed networks, connecting the radio telescopes through national research networks and 
the pan-European research network G\'EANT 2 via the Dutch national research network SURFnet to JIVE. 
At the time of our observations, a sustainable data rate of
128 Mbit s$^{-1}$ could be achieved. This is expected to improve
significantly in the near future. Description of some of the 
development work in this field was given by \cite{Par03,Szo04} 
and reports of more recent development are in preparation 
(Szomoru et al.; Strong et al.). The data were
transferred using Mark5A recorders with 1 Gbit s$^{-1}$ network
interface cards \citep{Whi03}.

We observed Cygnus X-3 at 5~GHz on three occasions, on 2006 March 16,
April 20 and May 18 using the six antennas forming the current
European e-VLBI network (e-EVN): Onsala 20~m, Torun, Jodrell Bank MkII, Cambridge,
Medicina and Westerbork (phased array). The observing run on March
16 was made in response to the first open call for e-EVN observations two weeks earlier. 
Unfortunately, due to technical problems with the off-line
correlator software, the data from March 16 were not useable and have
been omitted from the present study. In our second e-VLBI session, on
April 20, we observed Cygnus X-3 for an effective integration time of about 7
hours in a complex experiment consisting of interleaved observations
of Cygnus X-3 and GRS 1915+105 (see the accompanying paper -- Rushton
et al. (submitted) -- for the latter target). Occasionally,
delay and rate jumps were introduced in the data at the correlator Station Units, 
caused by a combination of limited memory buffer 
size and long correlation jobs. Although the effect was station based and thus the
closure phases were unaffected, this may have resulted in erroneous
phase transfer in a couple of phase-referencing scans. The problem
was eventually solved by self-calibrating the target data. During this
run, the target was in a `quiescent' state, with the flux density
levels in the decay phase following the major radio flares in 2006
March. At the beginning of May, Cygnus X-3 underwent another
significant flare and we observed it during an active state via a
target of opportunity proposal for around 12 hours on May 18 (Fig.~1). 
During both observations the amplitude of the flux density variations was 
around 10 percent, except for some $\sim$1~h long larger enhancements in the flux density registered 
at the beginning and end of the April run and at the beginning of the May run. Although 
this might have introduced artifacts in the radio images, we are confident that they 
would not be of significant importance.

The data were calibrated in \textsc{aips} (e.g.\ \citealt{Dia95})
using standard procedures. For phase-referencing we used J2007+4029, a
bright ($\sim$2~Jy correlated flux) calibrator located 4\fdg7 away from the target, 
employing a cycle time of 8 min (5 min on the 
target, 3 min on the phase-calibrator). The solutions derived from J2007+4029 
were smoothed, extrapolated and applied to Cygnus X-3. The a priori amplitude 
calibration was improved using the unresolved
 calibrator J2002+4725. Bandpass calibration was employed, using J2007+4029 in
epoch I (2006 April) and J2002+4725 in epoch II (2006 May). We used
J2007+4029 as fringe finder in the former run and OQ208 in the
latter. The data were self-calibrated once in phase with a 30-s
solution interval and once in amplitude and phase simultaneously with
a 30-min solution interval. The resulting radio maps are presented in
Fig.~2.

The instrumental polarisation leakage (the D-terms) was also
determined for epoch II from observations of the unpolarised
calibrator OQ208.  For calibration of the electric vector position
angles (EVPAs) at VLBI scales we analyzed separately the J2007+4029 
Westerbork Synthesis Radio Telescope (WSRT) data.
We found that this calibrator has a fairly compact polarisation
structure at VLBI scales enabling us to calibrate the EVPAs with an
estimated error of 10 degrees. The distribution of the EVPAs (in the
observer frame) and the fractional linear polarisation of Cygnus X-3
are shown in Fig.~3. The level of Faraday rotation due to the galactic
interstellar medium, affecting the overall distribution of EVPAs, is
unknown and therefore unaccounted for. For an arbitrary high rotation 
measure in the galactic plane \citep*{Sim81} of, say, 200 rad m$^{-2}$, the corresponding 
rotation of the EVPAs at 5~GHz would be up to $\sim$40\degr.

\begin{figure}
  \label{2006}
  \includegraphics*[angle=-90,scale=0.35]{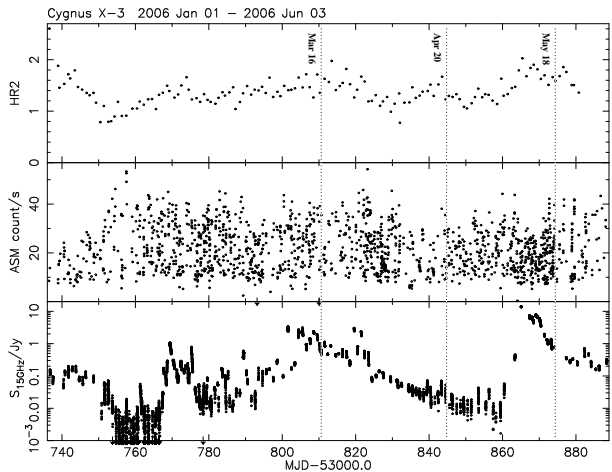}
  \caption{{\it{Top}}: {\it RXTE}/ASM hardness ratio HR2=5--12 keV / 3 -- 5 
keV. {\it{Middle}}: {\it RXTE}/ASM 2-10 keV X-ray light curve. {\it{Bottom}}: 15~GHz
radio (Ryle Telescope) light curve. Vertical lines indicate the
dates of our e-VLBI observations.}
\end{figure}

\begin{figure*}
  \includegraphics*[scale=0.35]{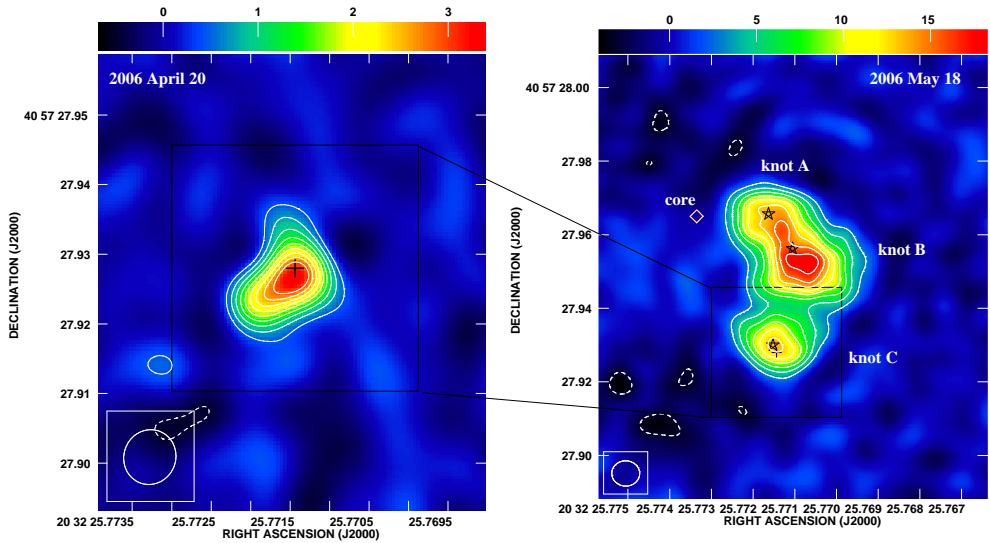}
  \caption{{\it{Left}}: 5~GHz e-VLBI radio map of Cygnus X-3 on 2006 April 
20 (epoch I). The contours are at -15, 15, 25, 35, 45, 55, 65,
75, 85 and 95 times the rms noise of 0.03 mJy beam$^{-1}$.  The beam
size is 7.9 $\times$ 7.4 mas$^2$ at PA=-19\fdg 6. The colour code bar
on top of the map is expressed in mJy beam$^{-1}$.  {\it{Right}}:
5~GHz e-VLBI radio map of Cygnus X-3 on 2006 May 18 (epoch II). The
contours are at -15, 15, 25, 35, 45, 55, 65, 75, 85 and 95 times the
rms noise of 0.2 mJy beam$^{-1}$. The beam size is 7.6 $\times$ 6.9
mas$^2$ at PA=88\fdg 3. The colour code bar on top of the map is
expressed in mJy beam$^{-1}$.  The positions of the centres of
the elliptical Gaussians fitted to the data are represented by a cross for epoch I 
and stars for epoch II.  The estimated location of the radio core from
Miller-Jones et al. (2004), J\,2000 coordinates: $ {\rmn{RA}} \
20^{\rmn{h}}32^{\rmn{m}}25\fs77335 $, $ {\rmn{Dec.}} \
40\degr57\arcmin27\farcs9650 $, is indicated with a diamond.  Note the change in
scale between the two images.} 
\end{figure*}

\section{Results and Discussion}

Fig.~2 presents the images from 2006 April 20 (epoch I) and May 18
(epoch II). Epoch I reveals a single faint component, possibly
slightly resolved, while epoch II reveals a bright curved structure
based around three bright knots. Also indicated on Fig.~2 is the
location of the X-ray binary radio core as estimated by Miller-Jones
et al. (2004; consistent also with Mioduszewski et al. 2001).

In order to facilitate comparison between these two epochs, and more
generally to allow us to extract as much information as possible from
the observations we fitted the data in the {\it uv}-plane with
elliptical Gaussians using \textsc{difmap} \citep{She97}.  For epoch I
we used one Gaussian while for epoch II we used as a starting model
three Gaussians at the positions of the bright features as seen in the
radio image before self-calibration. The results are summarized in
Table 1. The errors in the coordinates consist of the errors in the
position of the calibrator J2007+4029 ($\sigma_{\alpha}$=0.000180 s;
$\sigma_{\delta}$=0.00240 arcsec; \citealt{Fey04}) plus the error in
the phase-referencing procedure itself (estimated as 0.3 mas; \citealt*{Pra06}). 
We note that due to the poor {\it uv}-coverage at short baselines 
the fitted models are not well constrained, particularly the size of the Gaussians 
reported are probably
slightly underestimated.  Scatter-broadening is an issue for Cygnus
X-3; its geometric mean size is 15 $\pm$ 6 mas at 5~GHz
\citep*{Wil94,Sch95} and might be correlated with the flux at short 
VLBA baselines \citep*{New98}.

We draw attention to the fact that the maps shown in Fig.~2 account 
for a small fraction of the total flux density of the source. 
The 5~GHz e-VLBI integrated flux density measured in the
image plane for epoch I is 4.8~mJy. However, in the {\it uv}-plane,
the data were fitted well by a (larger) elliptical Gaussian with a 
flux density of 87 mJy.  For epoch II, the total flux density in the
{\it uv}-plane (determined by adding the flux densities contributed by
the three elliptical Gaussians fitted to the data) is 909~mJy (much 
of the flux being observed on the shortest baseline), while
in the image plane only 200~mJy is recovered. The explanation for
these differences resides in the fact that because of the lack of {\it
uv}-coverage at short baselines the large scale emission could not be
imaged properly. It is also worth mentioning that the simultaneous 
Westerbork integrated flux density in epoch II is 1.5 Jy, so at the
e-VLBI scale we are missing almost 40 per cent of the emission due to
the unavailability of very short baselines. As an overall comment, the
relatively sparse sampling at intermediate baselines has resulted in
some considerable loss of flux and uncertainty as to the true
strengths of individual knots. While we are confident that the
morphology of the imaged structure is real, it is clear that there is 
a significant, more diffuse component present which we were unable to image.

The single radio-emitting component from epoch I and knot C from epoch
II are positionally coincident within uncertainties.  This
identification of the radio emitting regions in quiescence and
outburst might intuitively lead to the conclusion that the core was
located here. However, the large positional discrepancy with the
previously estimated core position, plus the high degree of linear
polarisation (10 per cent, Fig.~3) in the region suggests that knot C
is probably not the core (see further discussion below). Therefore,
epoch I probably shows not the core but extended radio emission
physically separate from the X-ray binary which may be associated with
the 2006 March flaring events (Fig.~1).  This in turn implies that the
core itself was below detection levels, with a 3$\sigma$ upper limit
of 0.4\,mJy, about two orders of magnitude fainter than typical levels
for the source, and four orders of magnitude fainter than the
brightest flares.  Such `quenched' emission, lasting for periods of
weeks to months, was reported before other major flares of Cygnus X-3
\citep{Wal94,Wal95,Fen97}. 

\begin{figure*}
  \includegraphics*[scale=0.35]{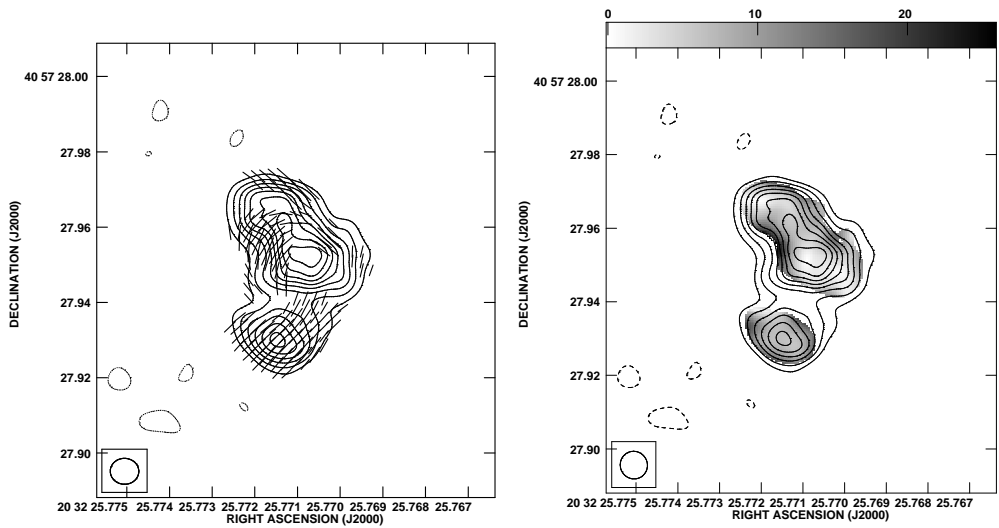}
  \caption{5~GHz e-VLBI radio map of Cygnus X-3 on 2006 May 18 (epoch II)
with the distribution of the EVPAs ({\it{left}}) and of the fractional
linear polarisation ({\it{right}}) i.e. the ratio between the
polarised flux and the total I flux, (Q$^2$+U$^2$)$^{1/2}$I$^{-1}$
with I, Q and U being the Stokes parameters. The code bar is expressed in 
percentage. In both images the Stokes I contours are at -15, 15, 25, 35, 45, 
55, 65, 75, 85 and 95 times the rms noise of 0.2 mJy beam$^{-1}$.}
\end{figure*}

\begin{table*}
 \centering
  \caption{Results of the {\it uv}-plane elliptical Gaussian fitting. The formal errors in coordinates (see text) are 0.000200 s in RA and 
0.00270 arcsec in Dec.}
  \begin{tabular}{lcccccc}
  \hline
   Epoch & Component & RA (J2000) & Dec.\ (J2000) & Size & PA & Flux density  \\
 & & & & (mas $\times$ mas) & (deg) & (mJy) \\
\hline
epoch I - April 20, 2006 & single & $20^{\rmn{h}}32^{\rmn{m}}25\fs771437$ & $+40\degr57\arcmin27\farcs92800$ & 17.5 $\pm$ 0.3 $\times$ 8.8 $\pm$ 0.5 & 150 
$\pm$ 2 & 87 $\pm$ 9\\
epoch II - May 18, 2006 & A & $20^{\rmn{h}}32^{\rmn{m}}25\fs771628$ & $+40\degr57\arcmin27\farcs96566$ & 23.2 $\pm$ 0.7 $\times$ 10.6 $\pm$ 2.5 & 
69 $\pm$ 13 & 60 $\pm$ 6 \\ 
                        & B & $20^{\rmn{h}}32^{\rmn{m}}25\fs771051$ & $+40\degr57\arcmin27\farcs95621$ & 38.4 $\pm$ 2.1 $\times$ 30.7 $\pm$ 8.1 & 
82 $\pm$ 22 & 748 $\pm$ 17 \\
                        & C & $20^{\rmn{h}}32^{\rmn{m}}25\fs771525$ & +40$\degr$57\arcmin27\farcs93000 & 19.4 $\pm$ 1.1 $\times$ 12.2 $\pm$ 2.3 & 
55 $\pm$ 7 & 101 $\pm$ 3 \\
\hline
\end{tabular}
\end{table*}

Epoch II reveals a much brighter and more complex structure, with the
appearance of a curved jet with three bright knots. Knot A is clearly
much closer to the previously estimated core location, but is still
$\sim$24 mas away. Mioduszewski et al. (2001) and Miller-Jones
et al. (2004) report the same core position to within 1 mas in data
separated by 4.5 years (but note that this is rather uncertain given
the fact that Mioduszewski et al. reported that their
phase-referencing was not entirely successful). For a distance to Cygnus X-3 of $\sim$9 
kpc \citep{Pre00} and an arbitrary speed in the plane of the sky of 
100 km s$^{-1}$, the upper limit to the positional shift due to the proper motion in the 
time interval between the observations of Miller-Jones et al. and ours is $\sim$10.5 mas. Higher 
speeds are definitely possible (for instance GRS 1915+105, at $\sim$11 kpc has a 
proper motion of $\sim$5.8 mas yr$^{-1}$; \citealt*{Dha00}), but given the present knowledge we cannot 
confidently identify the proper motion of the system as the cause of the shift. It seems likely
therefore that even knot A may not correspond to the core, but simply
to the nearest knot to it, indicating that at this epoch the core is
also quenched. The orientation of the knots with respect to the core
implies that all three are associated with the same side of the jet
(likely to be approaching, although the complex environment of Cygnus X-3
may strongly influence appearances), and were ejected in the sequence
C -- B -- A. Associating knot C with the large radio flare which
occurred about a week earlier would imply proper motions in the jet of
order 10 mas d$^{-1}$, consistent with those reported in Miller-Jones
et al. (2004). For a distance of $\sim$9 kpc, this corresponds to a
projected velocity of $\sim 0.5c$, and an intrinsic jet velocity at
least as large.

It is interesting to compare the `quenching' of the core radio
emission to that observed in the soft X-ray state of black hole
candidate X-ray binaries (e.g.\ \citealt{Fen99b}).  \cite{Gal03} have
reported a correlation between X-ray and radio luminosities in black
hole candidates in the hard X-ray state, below which the radio
emission falls dramatically in softer X-ray states.  Assuming a
distance of 9 kpc, the core radio emission of Cygnus X-3 in epoch I is
almost two orders of magnitude below the correlation, comparable to
the strongest limits on quenching found for other systems. Since
neutron stars do not appear to show such strong quenching in soft
X-ray states (\cite{Mig04} -- admittedly this is based on a very small
sample), this may be circumstantial evidence that Cygnus X-3 contains
a black hole (although the extreme nature of the environment in this
system might be responsible in some other way). However, the extreme
scattering of X-rays in the system is always going to make it hard to
relate the X-ray behaviour to that of the other black hole candidates.
It is noteworthy that in many cases `residual' radio emission, such as
that we believe we are imaging in epoch I, will limit the extent to
which observations on arcsec or coarser angular scales can measure the
true extent or rapidity of such quenching. As a result, VLBI
observations such as these are required to truly probe
rapidly-quenched radio states in bright jet sources.

In epoch II, the integrated flux densities from WSRT (1.5~Jy at 5~GHz) and the Ryle
Telescope (0.8~Jy at 15~GHz; Fig.~1) reveal a steep spectrum ($\propto
\nu^{-0.6}$), indicative of optically thin synchrotron
emission. However, since the map presented in Fig.~2 recovers $\leq
15$ per cent of this integrated flux density, we cannot immediately draw the
conclusion that the imaged structures are also therefore optically
thin synchrotron emission. Nevertheless, based on simple physical
arguments and comparison with other relativistic jet sources, this is
likely, and is consistent with the relatively high levels of
polarisation observed along the jet. As noted above, knot C in
particular shows a high level of polarisation, which supports the
interpretation that it is not the core. The highest fractional
polarisation, almost 25 per cent, appears to the east of knot B. This
is suggestive of an interaction site between the ejected matter and
the surrounding medium. The same conjecture can be made about the
region of relatively high polarisation, 15 per cent, in the north of
the complex, albeit with lower confidence due to the fact that
artifacts might appear at the edges of the emission. The coincidence
of emission at the location of knot C in both epochs may suggest that
this is the location of some standing shock in the flow or a site of
repeated jet-ISM interactions.

To summarize, a possible explanation for the radio emission detected
in the two experiments is the following: the radio emitting region of
epoch I is a jet-like structure probably ejected during the March
activity. Epoch II shows multiple radio features of what appears to be
a one-sided jet.  The new ejecta interact with the material ejected
during the previous outburst and generate the high polarisation on the
eastern edge of the radio structure and a slight bending of the jet
in this region towards the south-west (note that precession of the jet is 
likely at work -- \citealt{Mio01,Mil04}, 
thus complicating the interpretation). The core -- corresponding to the location of the
X-ray binary, and base of the relativistic jet -- is probably not
detected at either epoch, indicating a temporary suppression of jet
production at these times. It is important to bear in mind that the
jet is very close to the line of sight, at an angle of less than $\sim
15\degr$ \citep{Mio01,Mil04} -- this will cause an exaggeration of
apparent bends in the jet, and an underestimate of true jet velocities
(see more extensive discussions in Mioduszewski et al. 2001 and
Miller-Jones et al. 2004).

\section{Conclusions}

We have presented results on the X-ray binary Cygnus X-3 obtained in the 
frame-work of the first open call for European e-VLBI
observations. The two successful
experiments captured Cygnus X-3 initially in an almost quiescent state
after the outburst in 2006 March, and subsequently in an active state
a few days after the major flare in 2006 May. By the second epoch, a
bright, curved, relativistic jet is observed. The total intensity and
polarisation radio maps provide evidence for interaction of the jet
with the surrounding medium and, if our interpretation is correct,
with the material ejected in the previous active state. In the first
epoch, two weeks prior to the onset of the major flare of 2006 May,
the core is in a deeply radio-quenched state, with a $3\sigma$ upper
limit to the flux density of 0.4~mJy. The use of e-VLBI enabled very
quick access to the data, practically within one day from the end of
the runs, compared to sometimes weeks in the `traditional' VLBI
experiments. This facility for rapid response and analysis of VLBI
data is of paramount importance for the study of transient sources,
enabling decision making with respect to potential follow-up
observations (not only at radio wavelengths) to be much more
effective.

\section*{Acknowledgments}
We thank the JIVE staff R.M.~Campbell and C.~Reynolds for
their valuable contribution to the technical success of the
experiments.  The European VLBI Network is a joint facility of
European, Chinese, South African and other radio astronomy institutes
funded by their national research councils. The Ryle Telescope
is operated by the University of Cambridge. The X-ray data were
provided by the ASM/RXTE teams at MIT and at the RXTE SOF and GOF at
NASA's GSFC. e-VLBI developments in Europe are supported by the EC DG-INFSO funded 
Communication Network Development project, `EXPReS', Contract No. 02662.

\end{document}